\documentclass[12pt]{article}
\usepackage{amssymb,amsmath,epsfig}
\allowdisplaybreaks

\begin{document}

\title{Dynamics of Particles Around a Regular Black Hole with Nonlinear Electrodynamics }

\author{{Abdul Jawad}$^1$\thanks{jawadab181@yahoo.com;
 abduljawad@ciitlahore.edu.pk}, {Farhad
 Ali}$^{2}$\thanks{farhadmardan@gmail.com}, {Mubasher Jamil}$^{3}$ \thanks{mjamil@sns.nust.edu.pk}
 and\\ {Ujjal
 Debnath}$^{4}$\thanks{ujjaldebnath@gmail.com}\\
\small $^1$Department of Mathematics, COMSATS Institute of Information\\
\small Technology, Lahore-54000, Pakistan.\\
\small $^2$Department of Mathematics, Kohat University of Science and\\
\small Technology, Kohat, Pakistan.\\
\small $^{3}$Department of Mathematics, School of Natural Sciences (SNS),\\
\small National University of Sciences and Technology (NUST),\\
\small H-12, \small Islamabad, Pakistan.\\
\small $^{4}$Department of Mathematics, Indian Institute of Engineering\\
\small Science and Technology, Shibpur, Howrah-711 103, India.}

\date{}
\maketitle

\begin{abstract}
We investigate the dynamics of a charged particle being kicked off
from its circular orbit around a regular black hole by an incoming
massive particle in the presence of magnetic field. The resulting
escape velocity, escape energy and the effective potential is
analyzed. It is shown that the presence of even a very weak magnetic
field helps the charged particles in escaping the gravitational
field of the black hole. Moreover the effective force acting on the
particle visibly reduces with distance. Thus particle near the black
hole will experience higher effective force as compared to when it
is far away.
\end{abstract}
Center of mass energy. 
\newpage

\section{Introduction}

Dynamics of particles is an important and challenging topic in the
astrophysical studies of black holes. Partly because if the
particles are accreted by black holes than black holes gain mass,
and possibly black holes hurl particles away at high relativistic
speeds due to the angular momentum barrier. Moreover high energy
charged particles can also escape by the collision with other
particles from their stable orbits around black holes and move in
the electromagnetic fields under the Lorentz force law. Particles in
the surroundings of black holes are significantly influenced by the
strong gravitational pull.

The observational evidences \cite{Fro,Borm} indicate that the
magnetic field arises due to plasma which may exist in the
surrounding of black holes in the form of an accretion disk or a
charged gas cloud \cite{Mck,Dobb}. The relativistic motion of
charged particles in the accretion disk generates the magnetic
field. This magnetic field also leads to the generation of gigantic
jets along the magnetic axes. Since general relativity (GR) predicts
that all kinds of fields produce gravitational effects via curvature
of spacetime, it is assumed that the magnetic field is relativity
weak such that it does not affect the geometry of the black hole yet
it may affect the motion of charged particles \cite{Zna,Bland}. Due
to the presence of strong gravitational and electromagnetic fields,
the motion of charged particles is in general chaotic and
unpredictable. Moreover the corresponding orbits are unstable except
the innermost-stable-circular orbit.

If the collision of two particles occur near the event horizon of
black hole than high center of mass energy is produced.
Ba$\tilde{n}$ados, Silk and West (BSW) \cite{BSW} have proposed the
mechanism for the collision of two particles falling from rest at
infinity towards the Kerr black hole and determined that the center
of mass (CM) energy in the equatorial plane may be highest in the
case of a fast rotating black hole or the extremal black hole. Lake
\cite{Lake1} found that the CM energy of the particles at the inner
horizon of Kerr black hole. Further, Wei et al \cite{Wei}
investigated that the CM energy of the collision of the particles
around Kerr-Newmann black hole. A general review of the collision
mechanism is found in \cite{Harada}. The CM energy of the collision
of particles near the horizons of Kerr-Taub-NUT black hole
\cite{Liu}, Kerr-Newman-Taub-NUT black hole \cite{Zak},
Plebanski-Demianski black hole \cite{Sharif2}, charged dilaton black
hole \cite{Pradhan1}, cylindrical black hole \cite{Said} have been
investigated. The BSW effect has been studied for different black
hole
\cite{Jacob,Berti,Hus,Bambi,Tur,PU,Zas,Wei1,Grib,Pour1,P1,P3,G,Har}.

Recently, the dynamics of particles moving around weakly magnetized
black holes have been investigated. Motion of a charged particle
near weakly magnetized Schwarzschild black hole has been analyzed in
\cite{Frolo,Frolo1,Zah}. The chaotic motion of a charged particles
around Kerr black hole near magnetic field has also been
investigated in \cite{Nak,Taka,Preti,Kopa1,Kopa,Frolov,Ig}. The
Circular motion of charged particles around Reissner-Nordstrom black
hole has been analyzed by Pugliese et al \cite{Pug}. The particle
dynamics around Riessner-Nordstrom black hole with magnetic field
has been studied by Majeed et al \cite{Majeed}. The dynamics of a
charged particle around a slowly rotating Kerr black hole immersed
in magnetic field has been discussed by Hussain et al \cite{Huss}.
Also the dynamics of particles around a Schwarzschild-like black
hole in the presence of quintessence and magnetic field has also
been discussed by Jamil et al. \cite{Jamil00}. In the present work,
we want to investigate the dynamics of particles around a regular
black hole surrounded by external magnetic field.

The organization of the work is as follows: In section \textbf{2},
we review the regular black hole metric with non-linear
electromagnetic source and study the effective potential and energy
of particle. Section \textbf{3} deals with the dynamics of charged
particle. In section \textbf{4}, we investigate the center of mass
energy of two colliding particles. In section \textbf{5}, we find
the effective force and analyze the graphical representations.
Finally, the results and conclusions of the work are presented in
section \textbf{6}. We adopt the units $c=G=1$ and the metric signature $(+,-,-,-).$

\section{Regular Black Hole With Non-linear Electromagnetic Source}

It is well-known that linear or Maxwell electrodynamics leads to
curvature singularity and multiple horizons in
Reissner–-Nordstr$\ddot{o}$m spacetime. This problem of curvature
singularity gets fixed when quadratic (or generalized) Maxwell
tensor is used. Regular black holes are solutions of Einstein's
field equations that have horizon(s) and but no curvature
singularity. To avoid the singularity, Bardeen \cite{Bardeen}
proposed the concept of regular black hole, dubbed as Bardeen black
hole and subsequently, another type of regular black hole (Hayward
black hole) was discovered \cite{Hayward}. Another kind of regular
black hole is Ayon-Beato-Garcya (ABG) black hole \cite{Ayon}. The
regular black holes \cite{Ans,Balart} have the properties that their
metrics as well as their curvature invariants are regular
everywhere. This type of black holes violates the strong energy
condition somewhere in the spacetime; however, some of these
solutions satisfy the weak energy condition (WEC) everywhere
\cite{Eli,Zasla}. Bronnikov \cite{Bron} has even proved a theorem
which asserts that the existence of electrically charged, static,
spherically symmetric solutions with a regular center is forbidden,
while the existence of the solutions with magnetic charges is
feasible. In this connection, it is pertinent to study magnetically
charged regular black holes for particle acceleration and particle's
escape. Several regular black hole solutions have been found by
coupling gravity to nonlinear electrodynamics \cite{Ans,Balart}
theories.

A static spherically symmetric regular charged black hole is given by \cite{Balart}
\begin{equation}
ds^{2}=f(r)dt^{2}-\frac{dr^{2}}{f(r)}-r^2({d\theta}^{2}+\sin^{2}\theta{d\phi}^{2}),\label{1}
\end{equation}
where
\begin{align}
f(r)=1-\frac{2M}{r}\bigg(\frac{2}{\exp(\frac{q^2}{Mr})+1}\bigg).\label{2}
\end{align}
where $M$ and $q$ are respectively the mass and charge of the black hole. Notice that $r=0$ is no more a curvature singularity here while the horizon(s) exist at $f(r_h)=0$.  It is important to mention that in literature, a variety of static and non-static regular black holes have been discussed while we have chosen one of them to study particle dynamics. We expect that the particle dynamics around different static regular black holes as investigated here, should be generic.

In order to study the motion of particles in the background of a
regular black hole, we employ the Lagrangian dynamics and use the
following Lagrangian
\begin{equation}
\mathcal{L}=f(r)\dot{t}^{2}-\frac{\dot{r}^{2}}{f(r)}-r^2(\dot{\theta}^{2}+\sin^{2}\theta\dot{\phi}^{2}).\label{3}
\end{equation}
The overdot denotes derivative with respect to proper time. It is
obvious from Eq. (\ref{3}) that Lagrangian is independent of the
time $t$ and the $\phi$ coordinates explicitly but not of their
derivatives, hence it leads to corresponding symmetry generators
(also called Killing vectors). The Killing vector fields $\mathbf{X}$ under which
the spacetime (\ref{1}) remains invariant (i.e. $\mathbf{X}\mathcal
L=0$) are
\begin{eqnarray}\nonumber
\mathbf{X_0}&=&\frac{\partial}{\partial t}, \quad
\mathbf{X_1}=\frac{\partial}{\partial \phi}, \quad
\mathbf{X_2}=\cos\phi\frac{\partial}{\partial\theta}-\cot\theta\sin\phi\frac{\partial}{\partial\phi},
\\\label{4}
\mathbf{X_3}&=&\sin\phi\frac{\partial}{\partial\theta}+\cot\theta\cos\phi\frac{\partial}{\partial\phi}.
\end{eqnarray}
The conservation laws corresponding to the symmetries are given in
Table \textbf{1}. In this table, $E$, $L_z$, $L_1$ and $L_2$
indicate total energy, azimuthal angular momentum and angular
momenta, respectively. If the motion of particles is considered in
the equatorial plane than the $\mathbf{X_2}$ and $\mathbf{X_3}$
become irrelevant.
\begin{table}
\caption {Conservation Laws} \label{tab:title} \centering
\begin{small}
\begin{tabular}{|c|c|}
\hline Generator & First integrals\\
\hline $\mathbf{X_0}$ & $E=f(r)\dot{t}$\\
\hline$\mathbf{X_1}$& $-L_z=r^2\sin^2\theta\dot{\phi}$\\
\hline$ \mathbf{X_2}$ & $L_1=r^2\left(\cos\phi\dot{\theta} -\cot\theta\sin\phi\dot{\phi}\right)$\\
\hline $\mathbf{X_3}$ & $L_2=r^2\left
(\sin\phi\dot{\theta}+\cot\theta\cos\phi \dot{\phi} \right)$\\
\hline
\end{tabular}
\end{small}
\end{table}

The total specific angular momentum of the particle is defined as \cite{Fro}
\begin{align}
L^2\equiv r^4\dot{\theta}^2+\frac{L_z^2}{\sin^2\theta}=r^2v_\bot+\frac{L_z^2}{\sin^2\theta}
.\label{5}
\end{align}
By using the normalization condition for four-velocity and solving for
$\dot{r}^2$, we obtain
\begin{align}
\dot{r}^2=f^2(r)\dot{t}^2-f(r)\bigg(1+r^2\dot{\theta}^2+r^2\sin^2\theta\dot{\phi}^2\bigg).\label{6}
\end{align}
Due to spherically symmetry, all $\theta=$constant planes will be equivalent to the equatorial plane $\theta=\frac{\pi}{2}$, hence using the later value and the conservation laws we get
\begin{align}
\dot{r}^2=E^2-f(r)\bigg(1+\frac{L_z^2}{r^2}\bigg).\label{7}
\end{align}
For the particle moving in a circular orbit around the black hole, $\dot{r}=0$, Eq.(\ref{7}) implies
\begin{align}
E^2=f(r)\bigg(1+\frac{L_z^2}{r^2}\bigg)\equiv U_\text{eff}(r)\label{8},
\end{align}
where $U_\text{eff}$ is the effective potential. It suggests that at the horizon(s), the total energy and/or the effective potential will vanish. Particles with different energy and angular momentum will move under entirely different potential energy curves as are depicted below.

Differentiating the effective potential $U_\text{eff}$ with respect to $r$ and setting it to
zero, we obtain
\begin{align}\label{9}L^2_{z}=\frac{2r^2\bigg(q^2e^{\frac{q^2}{Mr}}
-Mre^{\frac{q^2}{Mr}}-Mr\bigg)}{r^2(1+e^{\frac{q^2}{Mr}})^2
+2\bigg(3Mre^{\frac{q^2}{Mr}}-q^2e^{\frac{q^2}{Mr}}+3Mr\bigg)}.
\end{align}
It represents the critical angular momentum that a particle carry in an orbit where the effective potential has an extremum (i.e. maximum or minimum). Hence, the value of $E^2$ takes the form
\begin{align}
E^2=\frac{r^3(e^{\frac{q^2}{Mr}}+1)^3-16M^2r(e^{\frac{q^2}{Mr}}+1)}
{r^3(e^{\frac{q^2}{Mr}}+1)^3+2r(e^{\frac{q^2}{Mr}}+1)(3Mr(e^{\frac{q^2}{Mr}}+1)
-q^2e^{\frac{q^2}{Mr}})}.\label{10}
\end{align}
After collision, the energy of the particle takes the form
\begin{align}
E_n^2=f(r)\bigg(1+\frac{(L_z+rv_{\perp})^2}{r^2}\bigg).\label{11}
\end{align}
For the particle to escape from the ISCO, we require the escape
energy to be greater after collision than the total energy before
collision. Observe the presence of an extra term $rv_{\perp}$  in
(\ref{11}) due to collision. Solving Eq.(\ref{11}) for $v_{\perp}$,
we have
\begin{align}
v_{\perp}=\sqrt{\frac{E_n^2-f(r)}{f(r)}}-\frac{L_z}{r}.\label{12}
\end{align}
which is the minimum escape velocity for general values of $E$, $f(r)$ and
$L_z$.

\section{Charge particle motion in the presence of magnetic field}

In this section, we adopt the formalism from  the literature
\cite{Zah,Jamil00}. The Lagrangian of a particle of mass $m$
carrying an electric charge $q$ takes the form
\begin{align}
\mathcal{L}=g_{\mu\nu}u^\mu
u^\nu+\frac{q}{m}A_\mu u^\mu,\label{13}
\end{align}
where $A_\mu$ is the four vector potential for the electromagnetic
field. Considering the presence of an axially symmetric magnetic
field with strength $B$ around black hole, we obtain new constants
of motions for the charged particle as
\begin{align}
E=f(r)\dot{t},\quad L_z=r^2\sin^2\theta(\dot{\phi}+B), \label{14}
\end{align}
Using the Eurler-Lagrange equation for $r$, we get
\begin{align}
\ddot{r}=\frac{1}{2}\bigg(f(r)(2r\dot{\theta}^2+2r\sin^2\theta\dot{\phi}^2)
-f'(r)f(r)\dot{t}^2-\frac{\dot{r}^2f'(r)}{f(r)}\bigg),\label{15}
\end{align}
where
\begin{align}
&f(r)=1-\frac{2M}{r}\bigg(\frac{2}{e^{\frac{q^2}{Mr}+1}}\bigg),\quad
\dot{t}=\frac{E}{f(r)},\quad
\dot{\phi}=\frac{L_z}{r^2\sin^2\theta}-B.\label{16}
\end{align}

Using the value of $\dot{t}$ and $\dot{\phi}$ from Eq.(\ref{16}) in
the Lagrangian (\ref{13}), we get the energy and corresponding
effective potential as follows
\begin{align}
&E^2=\dot{r}^2+r^2\dot{\theta}^2f(r)+f(r)\bigg(1+r^2\sin^2\theta\bigg(\frac{L_z}
{r^2\sin^2\theta}-B\bigg)^2\bigg),\\&
U_\text{eff}=f(r)\bigg(1+r^2\sin^2\theta\bigg(\frac{L_z}{r^2\sin^2\theta}
-B\bigg)^2\bigg).\label{17}
\end{align}
In order to integrate the dynamical equations, we need to make
these equations dimensionless by using the the following
transformations \cite{Huss,Jamil00,Hsu}
\begin{align}
&2M=r_d,\quad \sigma=\frac{\tau}{r_d}, \quad {\rho}=\frac{r}{r_d},\quad
l=\frac{L_z}{r_d},\quad b=Br_d,\quad q=\frac{\alpha}{r_d}.\label{18}
\end{align}
Using the transformation given in Eq.(\ref{18}), the equation of
motion, energy and effective potential acquire the form
\begin{eqnarray}\nonumber
\ddot{{\rho}}&=&\frac{1}{2}\bigg(f({\rho})(2{\rho}\dot{\theta}^2+2{\rho}\sin^2\theta\dot{\phi}^2)
-f'({\rho})f({\rho})\do{t}^2-\frac{\dot{{\rho}}^2f'({\rho})}{f({\rho})}\bigg),
\\\nonumber
E^2&=&\dot{{\rho}}^2+{\rho}^2\dot{\theta}^2f({\rho})+f({\rho})\bigg(1
+{\rho}^2\sin^2\theta\bigg(\frac{l}{{\rho}^2\sin^2\theta}-b\bigg)^2\bigg),\\\label{U}
U_\text{eff}&=&f({\rho})\bigg(1+{\rho}^2\sin^2\theta\bigg(\frac{l}{{\rho}^2\sin^2\theta}-b\bigg)^2\bigg).
\end{eqnarray}
\begin{figure}
\centering
\includegraphics[width=6cm]{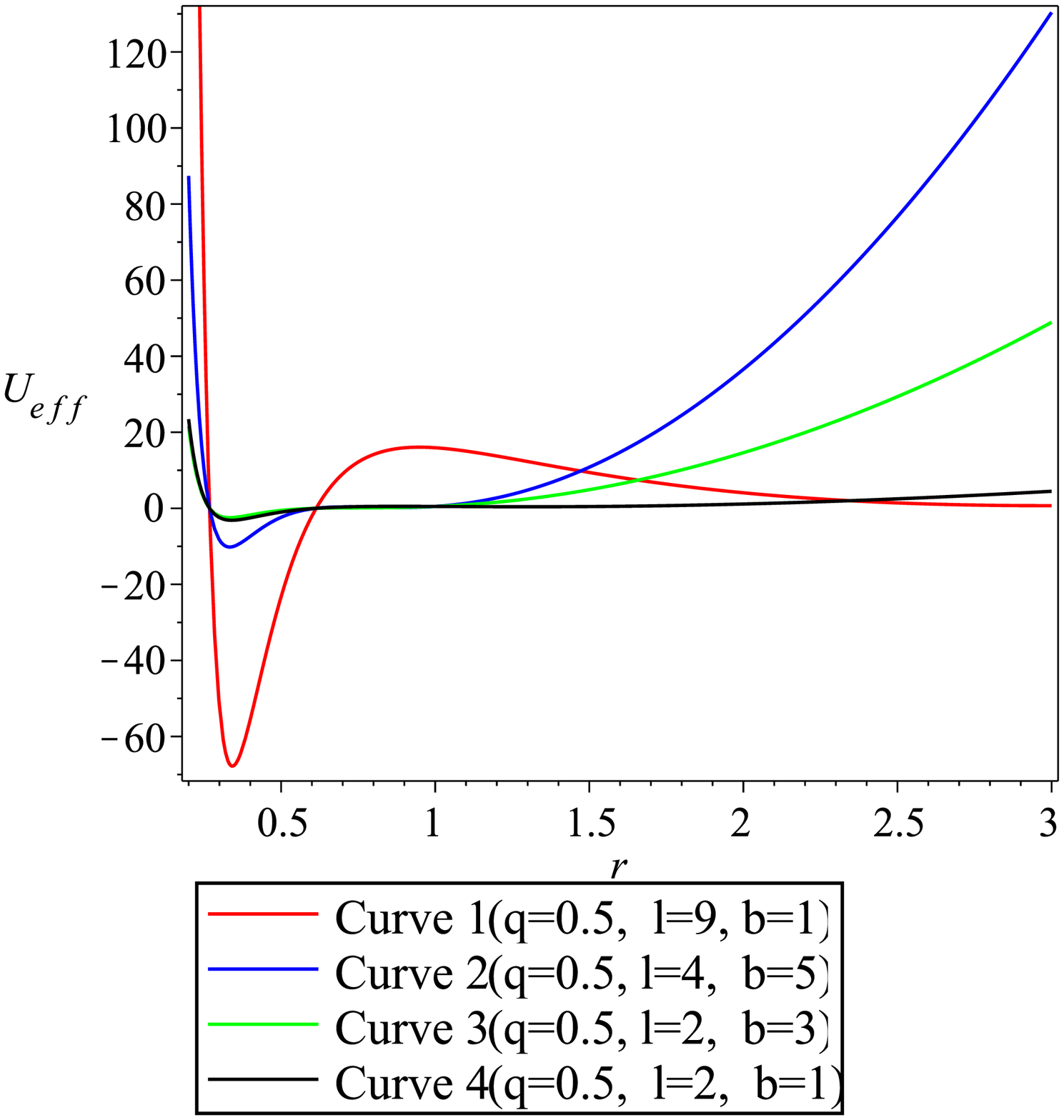}\includegraphics[width=6cm]{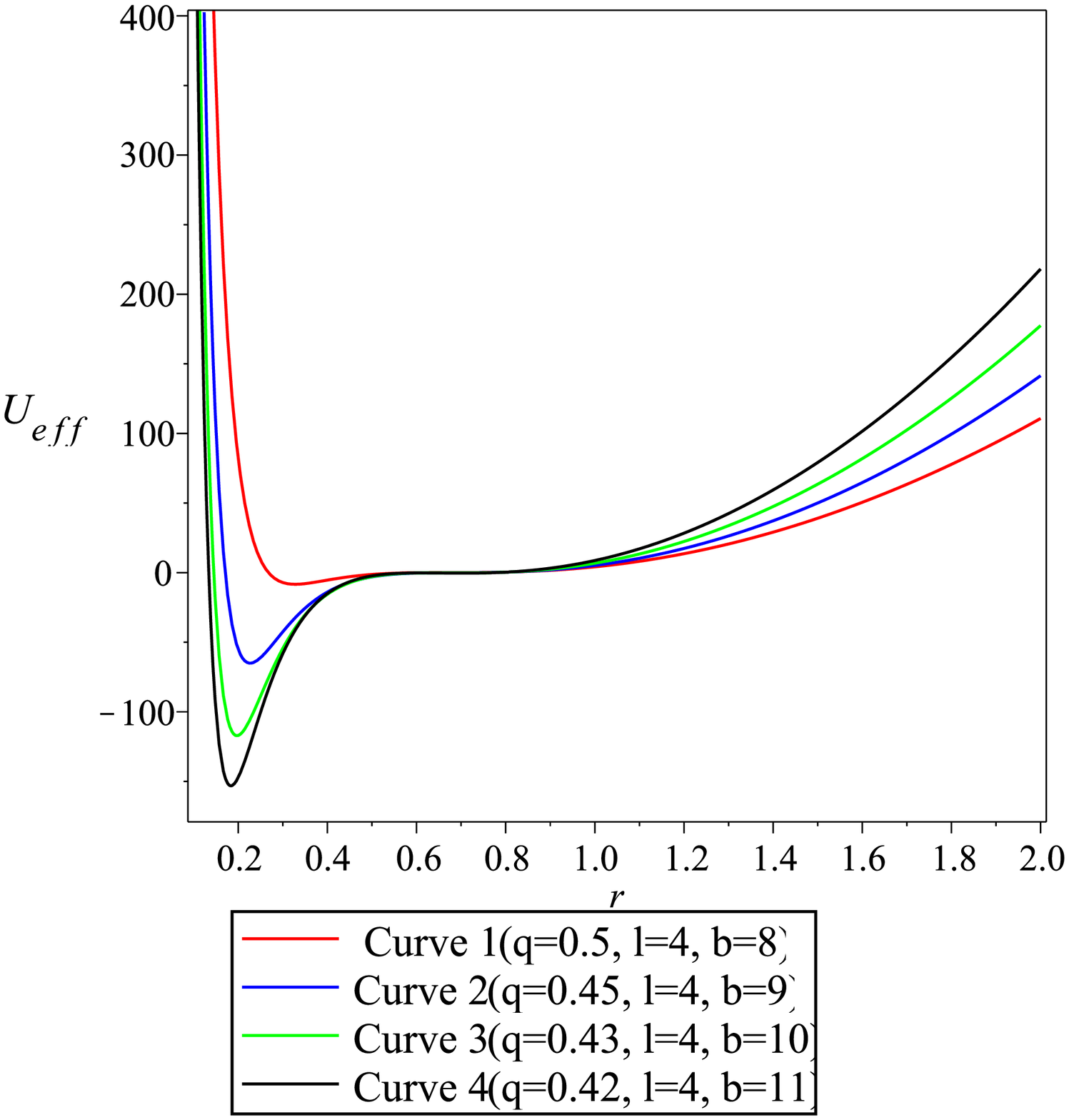}\\
\caption{The plot of effective potential ($U_\text{eff}$) versus $r$. In
the left panel, we fix $q=0.5$ and vary $l$ and $b$. In the right panel, we fix $l=4$ and vary $q$ and
$b$.}
\end{figure}
\begin{figure}
\centering
\includegraphics[width=6cm]{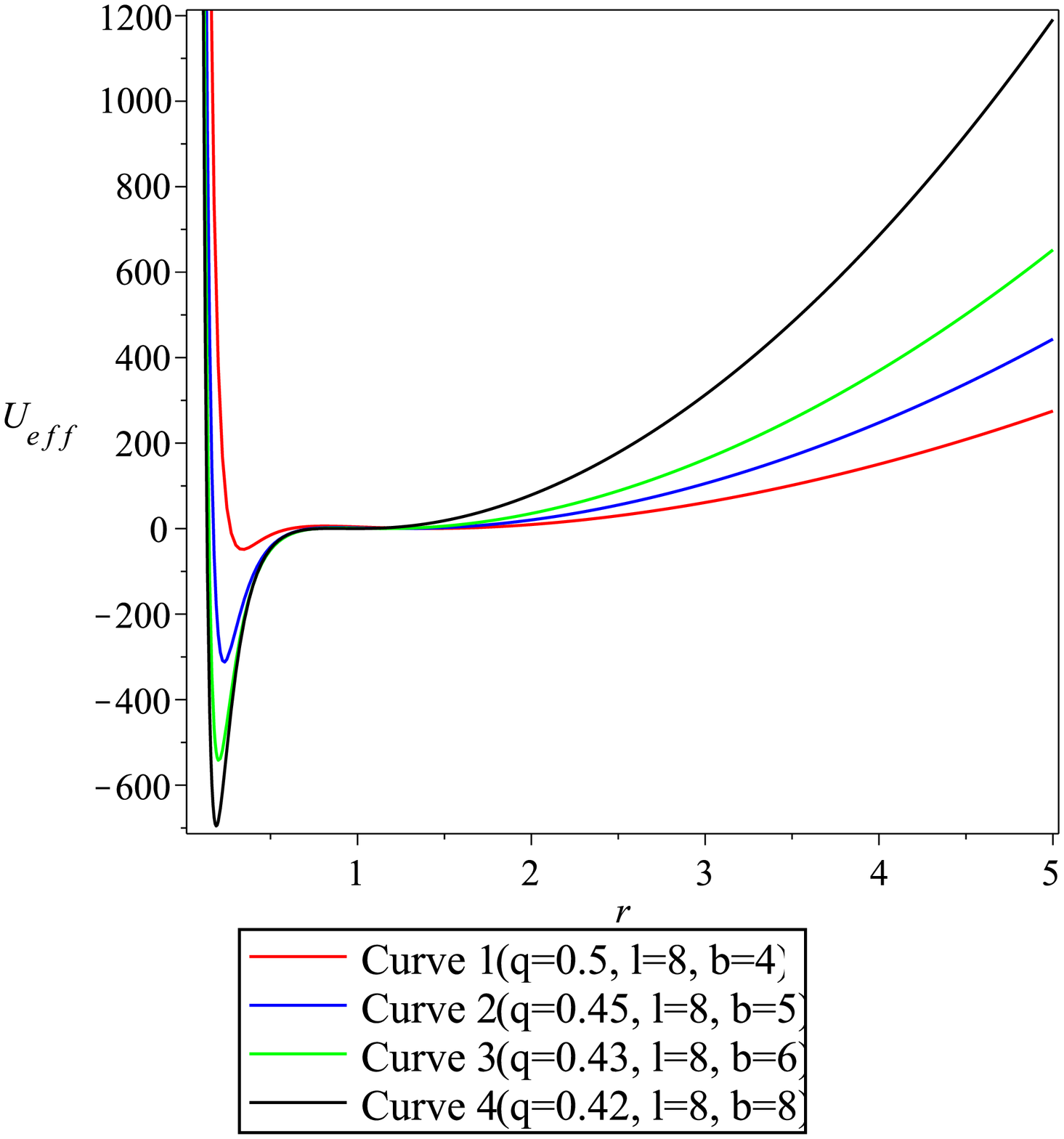}\includegraphics[width=6cm]{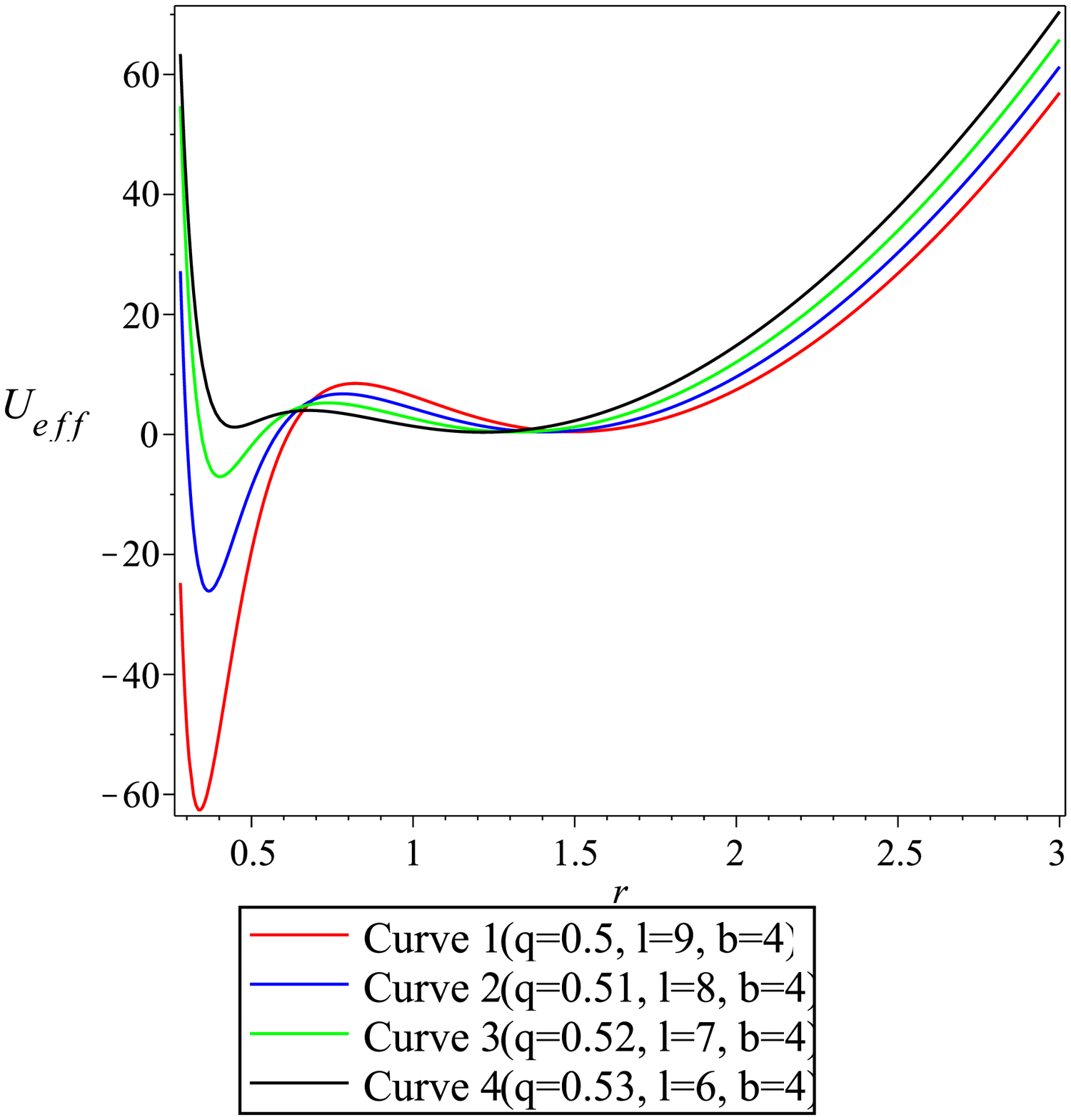}\\
\caption{The plot of effective potential ($U_\text{eff}$) versus $r$. In
the left panel, we fix $l=8$ and varying $q$ and $b$, respectively.
In the right panel, we fix $b=4$ and varying $q$ and $l$,
respectively.}
\end{figure}
We have drawn the graphs of effective potential $U_\text{eff}$ vs
radius $r$ for different values of $q,~l$ and $b$ in Figures
\textbf{1-2}. It is apparent that the effective potential has a
minimum near $r=0.2$ corresponding to a stable circular orbit.
However other circular orbits are also possible when the potential
has maxima.

Differentiating the energy of the particle with respect to $\rho$ by
taking $\theta=\frac{\pi}{2}$, we obtain
\begin{eqnarray}\nonumber
\frac{dU_\text{eff}}{d{\rho}}&=&\frac{2e^{\frac{2\alpha^2}{{\rho}}}(b^2{\rho}^5
+2\alpha^2b^2{\rho}^4+(1+2lb){\rho}^3-(2\alpha^2+4lb\alpha^2){\rho}^2-
3l^2{\rho})}{(e^{\frac{2\alpha^2}{{\rho}}}+1){\rho}^5}\\\label{20}
&+&\frac{2(b^2{\rho}^5+2
bl{\rho}^3-3l^2{\rho})}{(e^{\frac{2\alpha^2}{{\rho}}}+1){\rho}^5}+\frac{2(b^2l^2-l^2{\rho})}{{\rho}^4},
\end{eqnarray}
\begin{eqnarray}\nonumber
\frac{d^2U_\text{eff}}{d{\rho}^2}&=&-\frac{\bigg(\big(b^2{\rho}^5+2\alpha^2b^2{\rho}^4
+(1+2lb){\rho}^3-(2\alpha^2+4lb\alpha^2){\rho}^2
-3l^2{\rho}\big)\bigg)}{{\rho}^5(e^{\frac{2\alpha^2}{{\rho}}}+1)^2}\\\nonumber&\times&
\bigg(\frac{4\alpha^2e^{\frac{2\alpha^2}{{\rho}}}}{{\rho}^2}\bigg)+
\bigg(e^{\frac{2\alpha^2}{{\rho}}}
\big(5b^2{\rho}^4+8\alpha^2b^2{\rho}^3+3(1+2lb){\rho}^2-2(2\alpha^2\\\nonumber&+&
4lb\alpha^2){\rho}
-3l^2\big)\bigg)\bigg({\rho}^5(e^{\frac{2\alpha^2}{{\rho}}}+1)^2\bigg)^{-1}
+\bigg((e^{\frac{2\alpha^2}{{\rho}}}+1)^2(2l^5b^2-4
\\\nonumber&\times&
l^2{\rho})-\frac{8\alpha^2e^{\frac{2\alpha^2}{{\rho}}}}{{\rho}^2}(e^{\frac{2\alpha^2}{{\rho}}}+1)
(l^2b^2{\rho}-l^2{\rho}^2)\bigg)\bigg({\rho}^5(e^{\frac{2\alpha^2}{{\rho}}}+1)^2\bigg)-(\big(5
\\\nonumber&\times&{\rho}^4(e^{\frac{2\alpha^2}{{\rho}}}+1)^2-4{\rho}^3\alpha^2e^{\frac{2\alpha^2}{{\rho}}}
(e^{\frac{2\alpha^2}{{\rho}}}+1)\big)2e^{\frac{2\alpha^2}{{\rho}}}\big(b^2{\rho}^5
+2\alpha^2b^2{\rho}^4+{\rho}^3\\\nonumber&\times&(1+2lb)-(2\alpha^2+4lb\alpha^2){\rho}^2-3l^2{\rho}\big))({\rho}^{10}
(e^{\frac{2\alpha^2}{{\rho}}}+1)^4)^{-1}-\bigg(\big(5{\rho}^4\\\nonumber&\times&(e^{\frac{2\alpha^2}{{\rho}}}
+1)^2-4{\rho}^3\alpha^2e^{\frac{2\alpha^2}{{\rho}}}
(e^{\frac{2\alpha^2}{{\rho}}}+1)\big)(2l^5b^2-2l^2{\rho})\bigg)(
(e^{\frac{2\alpha^2}{{\rho}}}+1)^2\\\nonumber
&\times&{\rho}^{9})^{-1}-2\bigg(\big(5{\rho}^4(e^{\frac{2\alpha^2}{{\rho}}}+1)^2
-4{\rho}^3\alpha^2e^{\frac{2\alpha^2}{{\rho}}}
(e^{\frac{2\alpha^2}{{\rho}}}+1)\big)(2l^5b^2{\rho}^2+b^2\\\label{21}&\times&{\rho}^4-3l^2)\bigg)({\rho}^{9}
(e^{\frac{2\alpha^2}{{\rho}}}+1)^4)^{-1}+\frac{2(5b^2{\rho}^4+6bl{\rho}^2-3l^2)}{{\rho}^{5}
(e^{\frac{2\alpha^2}{{\rho}}}+1)^2}.
\end{eqnarray}
Here, $l$ and $b$ have these two relations given in Eqs.(\ref{20})
and (\ref{21}). After collision and for $\theta=\frac{\pi}{2}$ and
$\dot{{\rho}}=0$, the energy given in Eq.(\ref{U}) takes the form
\begin{align}
E^2=f({\rho})\bigg(1+{\rho}^2\bigg(\frac{l+{\rho}
v_\perp}{{\rho}^2}-b\bigg)^2\bigg).\label{22}
\end{align}
Solving equation (\ref{22}) for $v_\perp$, we get the escape
velocity for the charge particle as
\begin{align}
v_\perp=\frac{{\rho}\sqrt{E^2-f({\rho})}+b{\rho}^2-l}{{\rho}},\label{vp}
\end{align}
where
$f({\rho})=\bigg(1-\frac{1}{{\rho}}\bigg(\frac{2}{e^{\frac{2\alpha^2}{{\rho}}}+1}\bigg)\bigg)$.
We have drawn the graphs of escape velocity $v_\perp$ vs radius $r$
for different values of $E,~q,~l$ and $b$ in Figures \textbf{3-5}.
In left panel of Figure \textbf{3}, we are comparing the escape
velocity for different values of energy, magnetic field and angular
momentum by fixing the electric charge. We can observe that the
possibility of the particle to escape is higher for large values of
energy, magnetic field and angular momentum. In right panel of
Figure \textbf{4} , we are fixing the angular
 momentum and varying total energy, magnetic field and electric charge.
We can see that the behavior of escape velocity in Figure \textbf{4}
is somehow similar to Figure \textbf{3}. From Figure \textbf{4} and
\textbf{5}, we are observing the behavior of the particle to escape
from the vicinity of the black hole under the effect of magnetic
field. We can see from the figures that greater the strength of
magnetic field the possibility for a particle to escape is more.
Hence, we can conclude that the key role in the transfer mechanism
of energy is played by the magnetic field which is present in the
accretion disc to the particle for escape from the vicinity of black
hole. This is in agreement with the result of \cite{u1,u2}. However,
it is important to mention here that the trajectories of escape
velocity is entirely different from \cite{Majeed} due to presence of
exponential term in the regular blackhole metric.
\begin{figure}
\centering
\includegraphics[width=6cm]{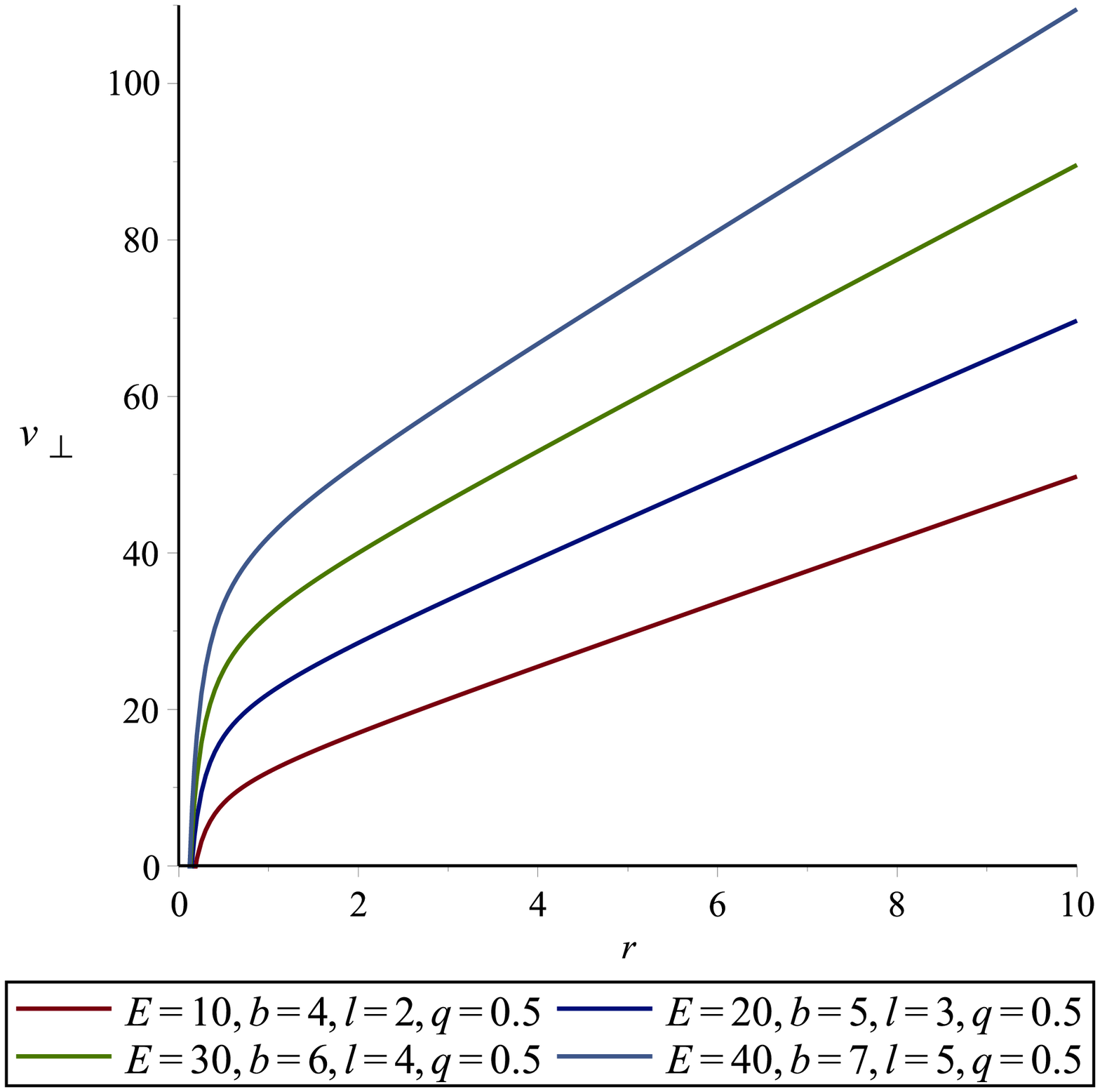}\includegraphics[width=6cm]{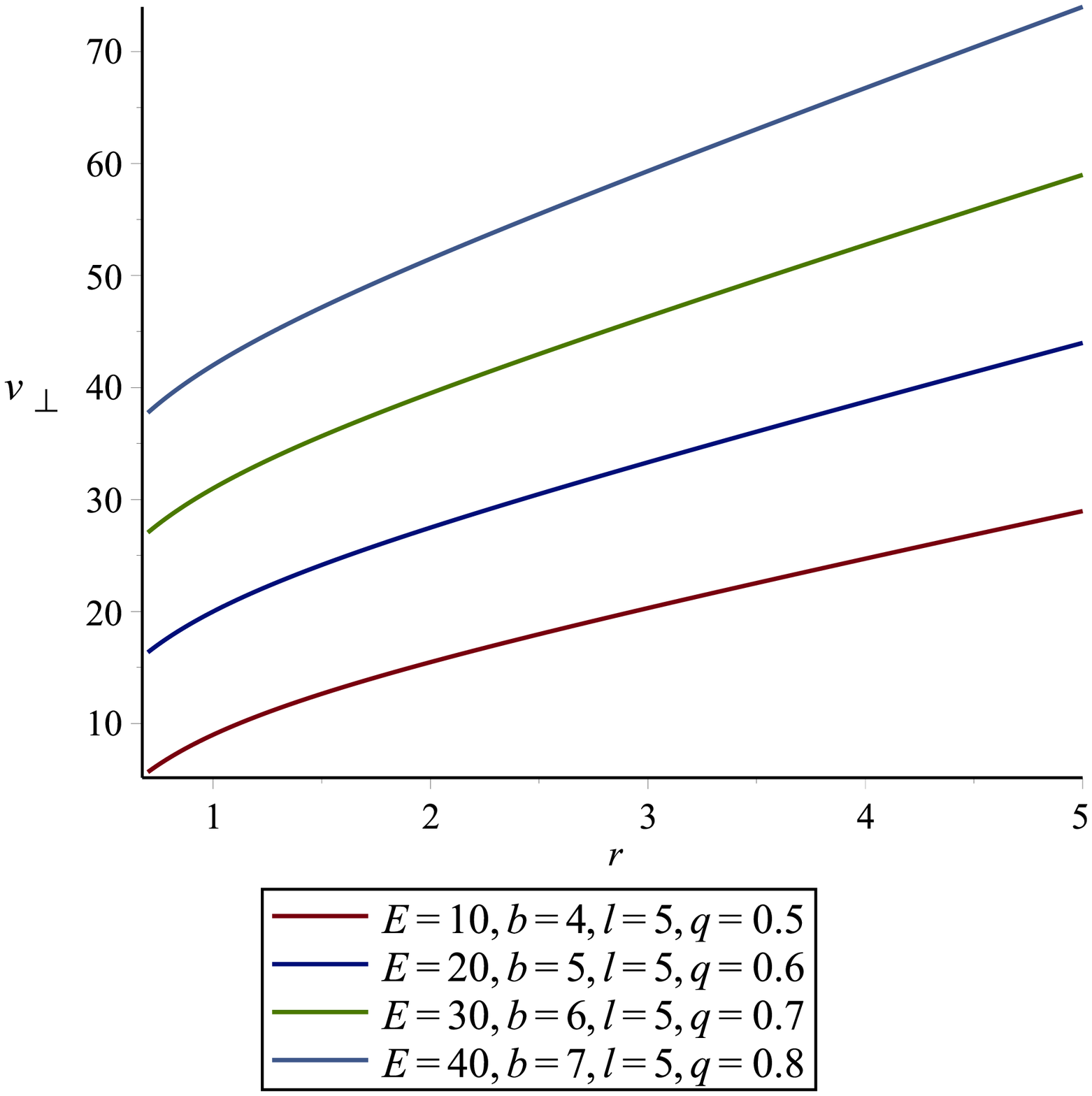}
\caption{The plot of escape velocity ($v_{\perp}$) versus $r$. In
the left panel, we fix $q=0.5$ and varying $l$ and $b$,
respectively. In the right panel, we fix $l=10$ and varying $q$ and
$b$, respectively.}
\end{figure}
\begin{figure}
\centering
\includegraphics[width=6cm]{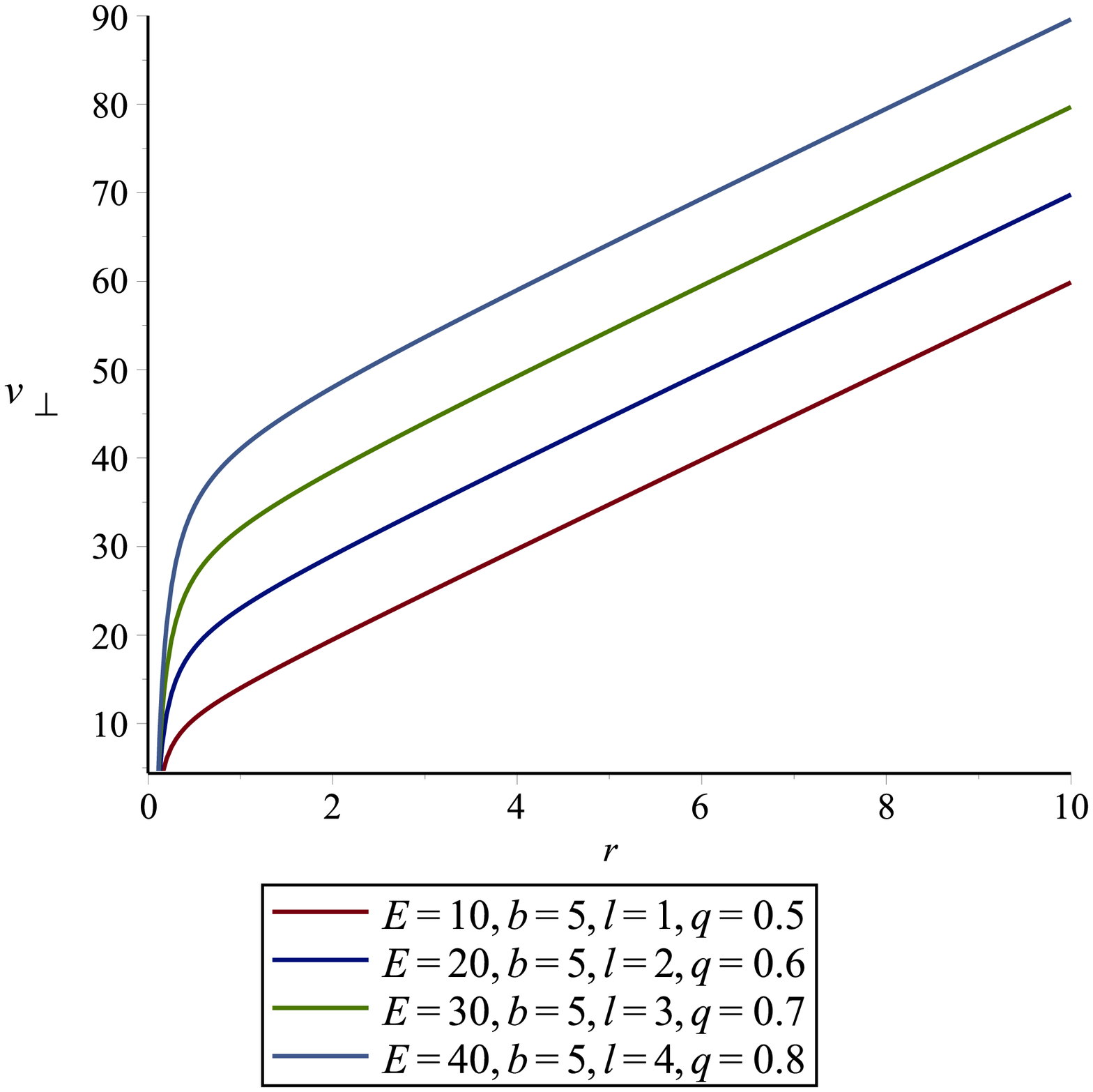}\includegraphics[width=6cm]{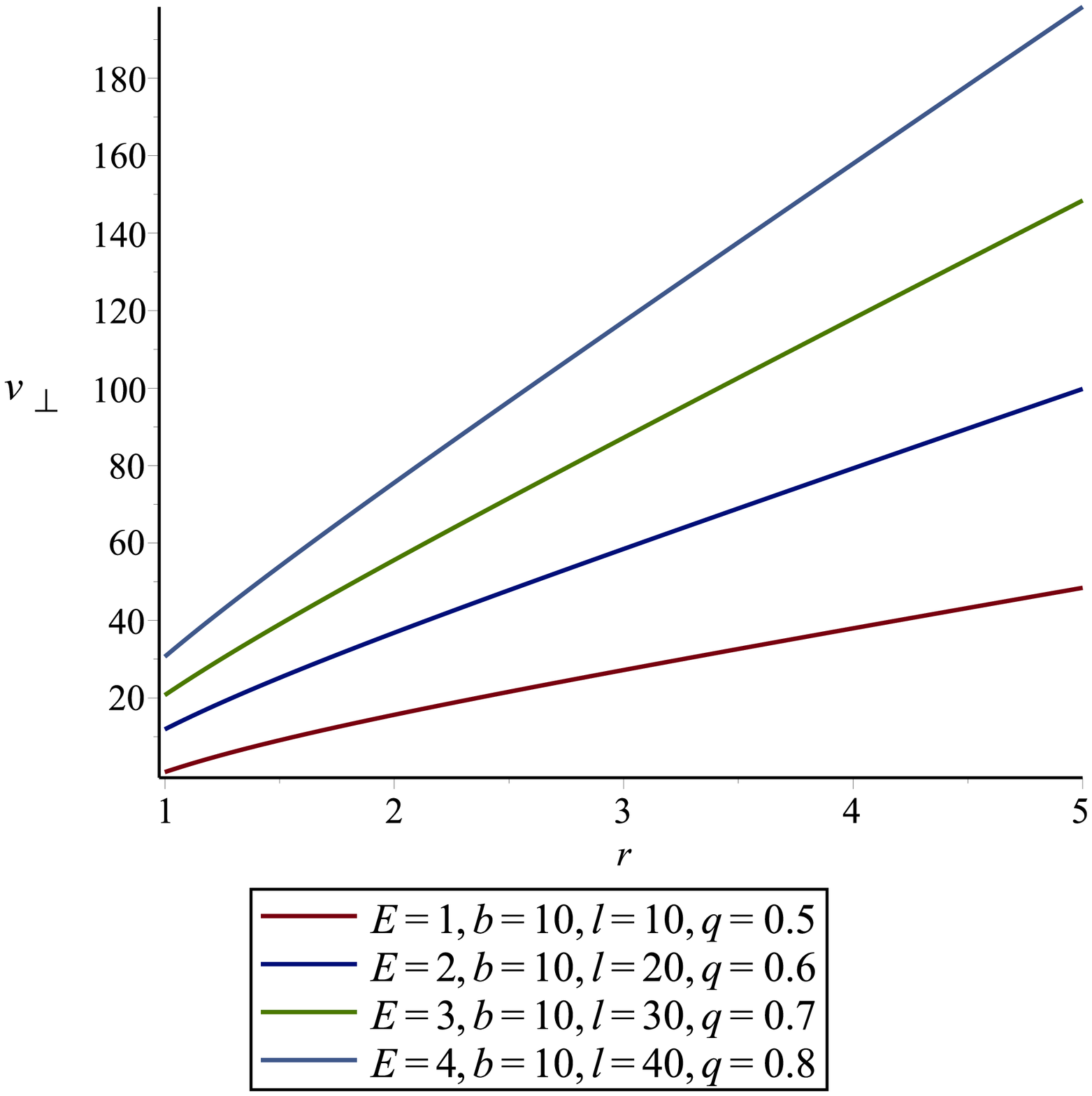}\\
\caption{The plot of escape velocity ($v_{\perp}$) versus $r$. In
the left panel, we fix $q=0.5$ and varying $l$ and $b$,
respectively. In the right panel, we fix $b=0$ and varying $q$ and
$l$, respectively.}
\end{figure}
\begin{figure}
\centering
\includegraphics[width=6cm]{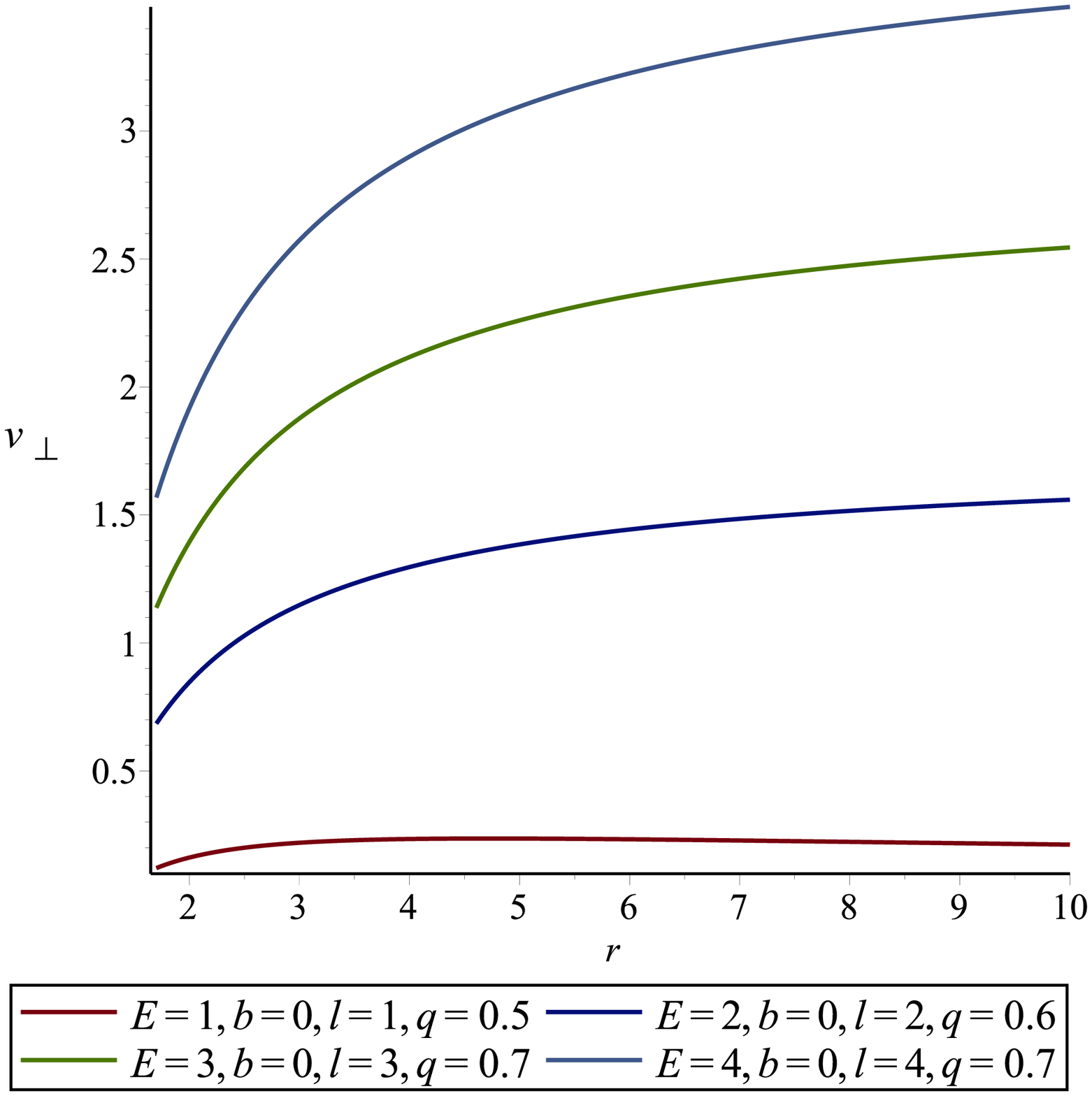}\\
\caption{The plot of escape velocity ($v_{\perp}$) versus $r$ by
fixing $b=10$ and varying $q$ and $l$, respectively.}
\end{figure}

\section{Center of mass energy of colliding particles}

\begin{itemize}
\item \textbf{Without magnetic field}:\\
The center of mass energy for colliding particle is define as
\begin{equation}
E_c=\sqrt{2}m_0(1-g_{\mu\nu}u^\mu u^\nu)^{\frac{1}{2}}.\label{24}
\end{equation}
where $u^\mu$ is the 4-velocity of each particles. Using the values
of Table \textbf{1} in Eq.(\ref{24}), we have
\begin{equation}
\begin{split}&E_c=\sqrt{2}m_0\bigg(2+\bigg(\frac{L_1^2+L_2^2}{2r^2}\bigg)
\bigg(\frac{E^2+f(r)}{E^2}\bigg)+L_1L_2\bigg(\frac{f(r)L_1L_2-2
r^2E^2}{2r^4E^2}\\\label{25}&+\frac{f(r)}{2E^2}\bigg)\bigg)^{\frac{1}{2}}.
\end{split}
\end{equation}
We want to obtain the CME near the horizon, where $f(r)=0$), and we
have
\begin{align}
E_c=2m_0\bigg(2+\frac{(L_1-L_2)^2}{4r^2_{h}}\bigg)^{\frac{1}{2}}.\label{26}
\end{align}

\item \textbf{With magnetic field:}

By the normalization condition
\begin{align}\dot{r}^2=f(r)\bigg(1+r^2+\bigg(\frac{L_z}{r^2}-B\bigg)^2\bigg),\label{27}
\end{align}
the center of mass energy takes the form
\begin{eqnarray}\nonumber
E_c&=&\sqrt{2}m_0\bigg(2+(L_1^2+L_2^2)\bigg(\frac{f(r)+E^2}{2E^2r^2}+\frac{B^2f(r)}{2E^2}\bigg)-(L_1^2+L_2^2)\\\nonumber&\times&
\bigg(\frac{(f(r)+E^2)B^2}{E^2}+\frac{B^3r^2f(r)}{E^2}\bigg)+L_1L_2\bigg((L_1L_2+2Br^2(L_1\\\nonumber&+&L_2)+4B^2r^4)\frac{f(r)}
{2E^2r^4}-\frac{1}{r^2}\bigg)+\bigg(B^4r^2(\frac{E^2+f(r)}{E^2})+\frac{f(r)}{2E^2}
\\\label{28}&+&\frac{f(r)r^4B^4}{4E^2}\bigg)\bigg)
^{\frac{1}{2}}.
\end{eqnarray}
The CME near horizon becomes
\begin{align}
E_c=2m_0\bigg(1+\frac{(L_1-L_2)^2}{4r_h^2}-\bigg({L_1+L_2-Br_h^2}\bigg)
\frac{B}{2}\bigg)^{\frac{1}{2}}.\label{29}\end{align}
\end{itemize}

\begin{figure}
\centering
\includegraphics[width=6cm]{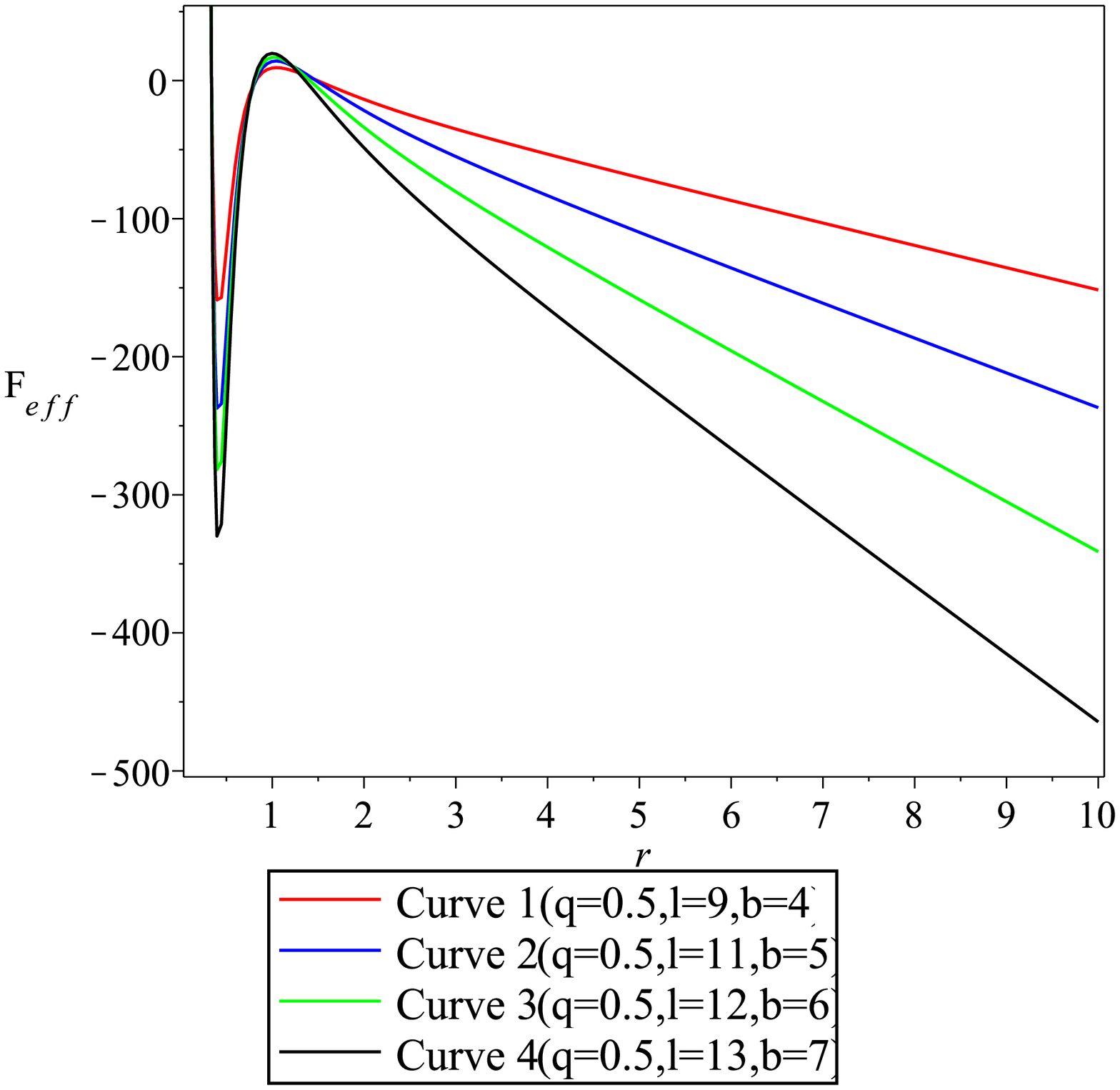}\includegraphics[width=6cm]{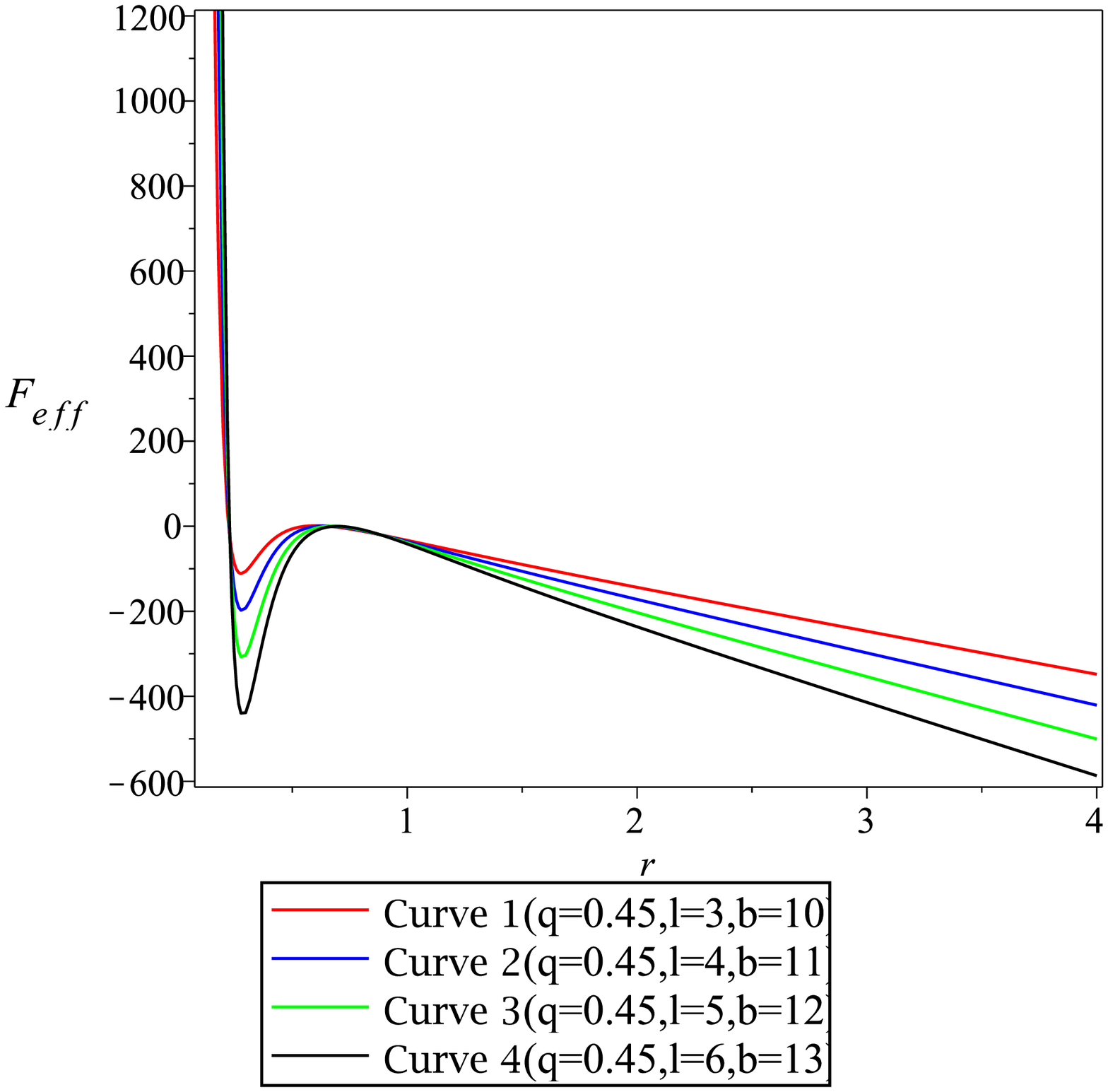}\\
\caption{The plot of effective force ($F_\text{eff}$) versus $r$ by
fixing $q=0.5$ (in the left panel) and $q=0.5$ (in the right panel)
while varying $l$ and $b$.}
\end{figure}
\begin{figure}
\centering
\includegraphics[width=6cm]{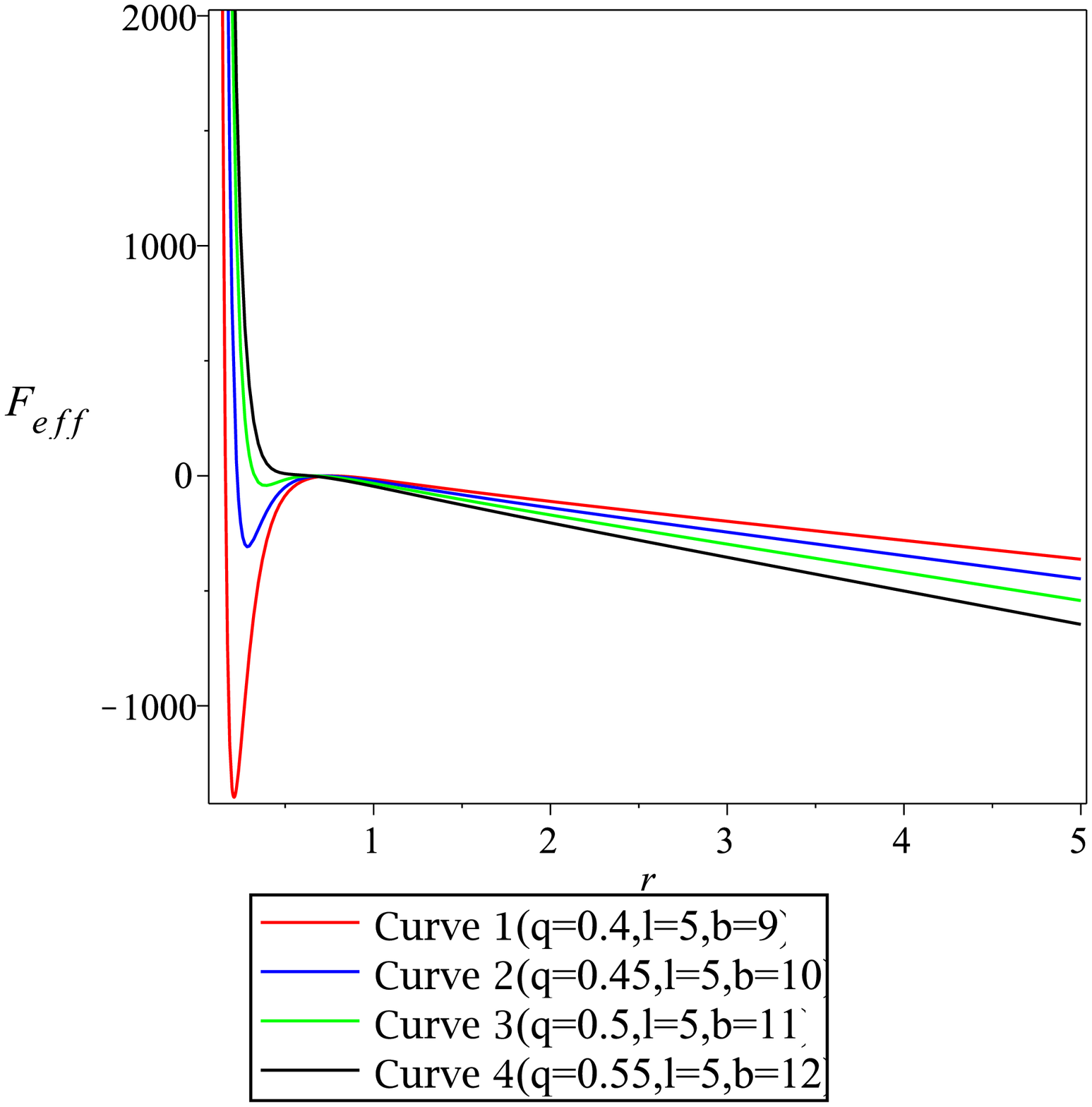}\includegraphics[width=6cm]{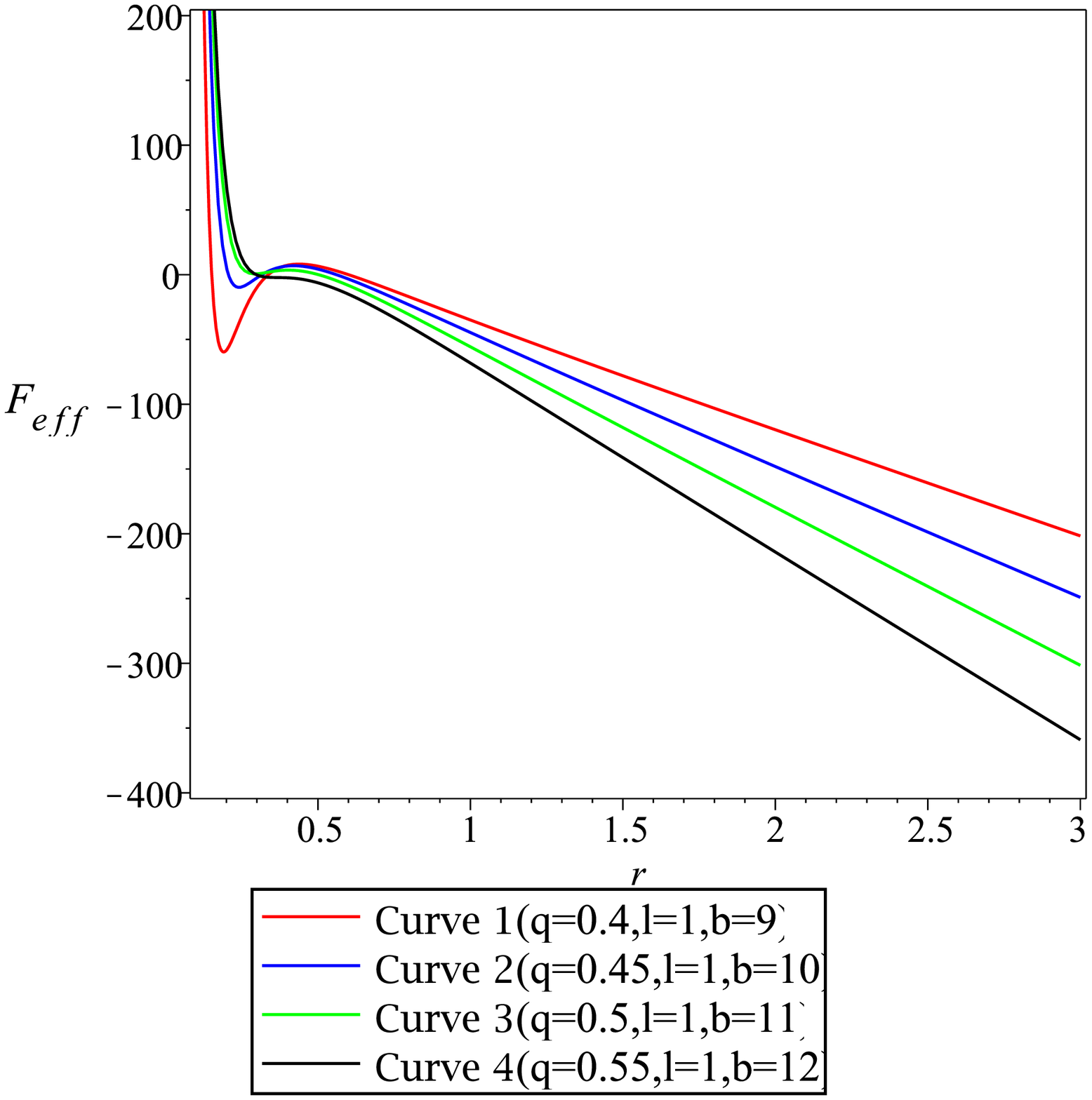}\\
\caption{The plot of effective force ($F_\text{eff}$) versus $r$ by
fixing $l=5$ (in the left panel) and $l=1$ (in the right panel)
while varying $q$ and $b$.}
\end{figure}
\begin{figure}
\centering
\includegraphics[width=6cm]{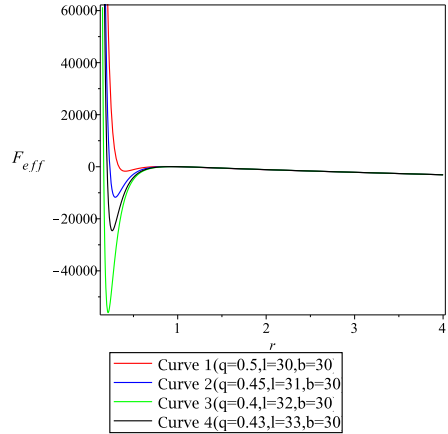}\\
\caption{The plot of effective force ($F_\text{eff}$) versus $r$ by
fixing $b=30$ and varying $q$ and $l$.}
\end{figure}

\section{Effective force}

For a particle moving in the electromagnetic field in a flat
background, the effective force is determined by the Lorentz force.
In a curved background, the same force acting on the particle can be
obtained by the derivative of effective potential as \cite{Jamil00}
\begin{eqnarray}\nonumber
F_\text{eff}=-\frac{1}{2}\frac{dU_\text{eff}}{dr}&=&-e^{\frac{2\alpha^2}{r}}(b^2r^5+2\alpha^2b^2r^4+(1+2lb)r^3-
(2\alpha^2+4lb\alpha^2)r^2\\\nonumber&-&3l^2r)((e^{\frac{2\alpha^2}{r}}+1)r^5)^{-1}-\frac{(b^2r^5+2blr^3-3l^2r)}
{(e^{\frac{2\alpha^2}{r}}+1)r^5}-(b^2l^2\\\label{f}&-&l^2r)r^{-4}.
\end{eqnarray}
For rotational angular variable \cite{Jamil00}, we obtain
\begin{align}
\frac{d\phi}{d\tau}=\frac{L_z}{r^2}-b\label{31}
\end{align}
The Lorentz force on the particle is repulsive if $L_{z}>br^{2}$ and
the Lorentz force is attractive if $L_{z}<br^{2}$. We have drawn the
graphs of effective force $F_\text{eff}$ vs radius $r$ for different
values of $\alpha,~l$ and $b$ in Figures \textbf{6-8}. It is
apparent from these figures that the effective force acting on the
charged particles will be small as the distance from the black hole
increases and vice versa if the distance shrinks. Thus at large
distance, black hole's gravity and the electromagnetic field will
have diminishing effects on the particles.

\section{Concluding Remarks}

In this work, first we have considered the regular black hole
metric with non-linear electromagnetic source and studied the
effective potential, escape velocity and energy of charged
particle in presence of magnetic field. Next, we have investigated
the center of mass energy of two colliding particles with and
without magnetic field.  Finally we found the
effective force and analyzed the graphical representations of
effective potential, escape velocity and effective force.

Although the black hole is regular (non-singular at all spacetime
points), the geometry of spacetime outside the black hole's horizon
should be similar to that of conventional black holes. Moreover the
particles moving around such black holes should not feel the absence
of the singularity. The present study also confirms the same
features. Charged particle after collision can escape if the
effective force acting on it is weaker. Hence, we can conclude that
the key role in the transfer mechanism of energy is played by the
magnetic field which is present in the accretion disc to the
particle for escape from the vicinity of black hole. This is in
agreement with the result of \cite{u1,u2}. However, it is important
to mention here that the trajectories of escape velocity is entirely
different from \cite{Majeed} due to presence of exponential term in
the regular black hole metric.

\end{document}